\documentstyle[12pt,aasms4,flushrt]{article}

\def\etal{et\ al.}
\def\CIV{C~{\sc iv}}
\def\MgII{Mg~{\sc ii}}
\def\hMpc{$h^{-1}$~Mpc}
 
\lefthead{Quashnock, Vanden Berk and York}
\righthead{High Redshift Superclustering}

\begin{document}

\title{High--Redshift Superclustering of QSO Absorption Line Systems\\
	 on 100 \hMpc\ Scales}

\author{Jean M. Quashnock\altaffilmark{1}, Daniel E. Vanden Berk, 
	and Donald G. York}

\affil{University of Chicago\\ Dept. of Astronomy and Astrophysics\\
 5640 S. Ellis, Chicago, IL 60637}
\altaffiltext{1}{{\it Compton} GRO Fellow -- NASA grant GRO/PDP 93-08.}

\authoremail{jmq@oddjob.uchicago.edu}
\authoremail{devb@oddjob.uchicago.edu}
\authoremail{don@oddjob.uchicago.edu}

\received{25 July 1996}
\slugcomment{To appear in {\it The Astrophysical Journal Letters}}

\bigskip
\noindent
\centerline{Electronic mail: jmq@oddjob.uchicago.edu ,
devb@oddjob.uchicago.edu ,}
\centerline{don@oddjob.uchicago.edu }

\begin{abstract}
We have analyzed the clustering of \CIV\ absorption line systems
in an extensive new catalog of heavy element QSO absorbers.
The catalog permits exploration of clustering
over a large range in both scale (from about 1 to over 300 \hMpc)
{\it and} redshift ($z$ from 1.2 to 4.5).
We find significant evidence ($5.0\, \sigma$; $Q=2.9\times10^{-7}$)
that \CIV\ absorbers are clustered
on comoving scales of 100 \hMpc\ ($q_0=0.5$) and less
--- similar to the size of voids and
walls found in galaxy redshift surveys of the local universe ($z < 0.2$) ---
with a mean correlation function 
$\langle\xi_{\rm aa}\rangle = 0.42 \pm 0.10$ over these scales.
We find, on these scales,
that the mean correlation function at 
low ($\langle z\rangle_{\rm low} = 1.7$),
medium ($\langle z\rangle_{\rm med} = 2.4$),
and high redshift ($\langle z\rangle_{\rm high} = 3.0$) is
$\langle\xi_{\rm aa}\rangle = 0.40 \pm 0.17$,
$0.32 \pm 0.14$, and
$0.72 \pm 0.25$, respectively.
Thus, the superclustering is present
even at high redshift;
furthermore, it does not appear that the superclustering scale,
in comoving coordinates, has changed significantly since then.
We find 7 QSOs with rich groups of absorbers 
(potential  superclusters) that account for a
significant portion of the clustering signal, with
2 at redshift $z\sim 2.8$.
We find that the superclustering is just
as evident if we take $q_0=0.1$ instead of 0.5;
however, the inferred scale of clustering is then 240 \hMpc ,
which is larger than the largest scales of clustering known at present.
This discrepancy may be indicative of
a larger value of $q_0$, and hence $\Omega_0$.
The evolution of the correlation function 
on 50 \hMpc\ scales is consistent
with that expected in cosmologies with
density parameter ranging from $\Omega_0= $ 0.1 to 1.
Finally, we find no evidence 
for clustering on scales greater than 100 \hMpc\
($q_0=0.5$) or 240 \hMpc\ ($q_0=0.1$).
\end{abstract}


\section{Introduction}

It has been recognized for some time now that 
QSO absorption line systems 
are particularly effective probes of large--scale structure
in the universe (see, e.g., \cite{Shav83,CMY85}).
This is because the absorbers trace matter lying on the QSO line of sight,
which can extend over a sizable redshift interval out to high redshifts.
Thus, the absorbers trace both the large--scale structure (on scales
out to hundreds of Mpc) {\it and} its evolution in time,
since the clustering pattern can be examined as a function of
redshift out to $z\sim 4$.
The evolution of large--scale structure is of great interest, since,
in the gravitational instability picture,
it depends sensitively on the mean mass density $\Omega_0$ 
(\cite{Peeb80}, 1993).

We are not concerned here with the relationship between the absorbers
and galaxy haloes, or with small--scale structure.
However, in another paper (\cite{Quash96}) we discuss
the small--scale clustering of absorbers and relate it to
galaxy clustering.
 
In this Letter, we present results of an analysis of
line--of--sight correlations of \CIV\ absorption line systems,
using a new and extensive catalog of absorbers (\cite{Y96}).
This catalog contains data on all QSO
heavy--element absorption lines in the literature, 
complete up to December 1995, 
with some additional entries since then.  It is an updated version of
the catalog of York \etal\ (1991), 
but is more than twice the size, with over 2200
absorbers listed over 500 QSOs, and is the largest sample of heavy--element
absorbers compiled to date.  More details can be
found in the earlier version of the catalog,
as well as in our recent paper (\cite{Vanden96})
in which we find that
the number of absorbers is correlated
with the intrinsic brightness of the QSO,
suggesting that QSOs are lensed by matter associated with the absorbers.
A preliminary study suggests that the absorber correlation function 
is not strongly dependent on the intrinsic brightness of the QSO.
We describe the correlation analysis in \S 2,
present our results in \S 3, and discuss the implications of these in \S 4.

\section{Correlation Analysis of \CIV\ Absorbers}

Here we briefly describe our procedure for calculating the 
line--of--sight correlation function, $\xi_{\rm aa}$.
Unless otherwise noted, we take $q_0=0.5$ and $\Lambda=0$.
We follow the usual convention and take $h$ to be the Hubble constant
in units of 100 km~s$^{-1}$~Mpc$^{-1}$.

To produce a more uniform set of absorbers from the inhomogeneous catalog,
we have applied the following selection 
criteria to the QSO spectra and absorbers.  
The Ly$\alpha$ forest region of each spectrum was
excluded because identification of heavy--element lines there is
problematic.
This means we examine absorbers within about 60,000 km~s$^{-1}$,
or about 400 \hMpc , of the QSO.
The so--called associated region within 5000 km s$^{-1}$
of each QSO emission redshift has also been excluded, 
since the number density of absorbers there 
may not be representative of that for absorbers
farther removed in redshift from their QSOs (\cite{Ald94}).   
Thus a typical redshift range for a line of sight is $\Delta z=0.4$.
All of the selected
absorption systems must at least have an identified \CIV\ doublet,
and if no other line was identified, the
equivalent width ratio of the 1548 {\AA} component to the
1551 {\AA} component must be $\geq 1$
within the listed measurement errors.  
In addition, the equivalent width of
the 1548 {\AA} component must have been detected at more than the 
$5\,\sigma$ level, and have a rest value of at least $0.1$ {\AA}.  
In order to avoid aliasing of power (\cite{Tytler93})
on the large scales of interest 
because of the very pronounced peak in the  correlation function  
on small scales (\cite{SBS,Quash96}),
we have combined all absorbers lying within 3.5 comoving \hMpc\
of each other ($\sim 600$ km s$^{-1}$)
into a single system with redshift equal to the average redshift
of the components
and with equivalent width equal to that of the strongest component.  
Because of this, we have included only spectra with resolutions
that allow absorber comoving separations $\leq 3.5$ \hMpc\ to be resolved.  

These selection criteria leave a total of 360 \CIV\
absorption systems, with redshifts from 1.2 to 4.5, scattered among
373 QSO lines of sight.
The distribution in the total number of absorbers
over the line of sight
is consistent with that from a uniform distribution.
For number of absorbers per line of sight 0 through 7,
the observed versus expected number of lines of sight are
(207, 190.34), (67, 82.52), (49, 52.23), (22, 28.24), (16, 12.92),
(9, 4.61), (1, 1.60), and (2, 0.39).
A KS--test shows that the two distributions are 
consistent with each other ($Q=0.45$).
This shows that the average distribution of absorbers
on 400 \hMpc\ scales is uniform.

To study the distribution on smaller scales,
we use the \CIV\ line--of--sight correlation function, 
$\xi_{\rm aa}$, which is 
calculated by binning the comoving separation, $\Delta r$,
of absorber pairs and comparing the number of real pairs, $N_r$, in a bin 
to the average number of pairs, $\overline N_s$,
generated from 100 Monte Carlo simulations; namely, $\xi_{\rm aa}(\Delta r)
= (N_{r}/\overline{N}_{s}) - 1$.
The simulated absorber samples are generated by randomly
redistributing all of the real absorbers among the QSO spectra,
subject to the condition that the absorbers 
could have been detected in both redshift
and equivalent width in their randomly selected QSO spectra.  This
technique is possible because the catalog of absorbers lists the wavelength
limits and equivalent width limits for each of the listed QSO spectra.
This technique has the virtue that the redshift and
equivalent width number densities of all the simulated data sets are
exactly the same as for the real data set, 
so that the real and simulated data sets
have the same evolution of physical properties as a function of redshift.
The $1\, \sigma$ region of scatter around the
null hypothesis of no clustering ($\xi_{\rm aa}=0$)
is given by the standard deviation of $\xi_{\rm aa}$ 
in the 100 random simulations,
whereas the $1\, \sigma$ error in the estimator of $\xi_{aa}$ in
each $\Delta r$ bin is $\sqrt{N_{r}}/\overline{N}_{s}$ (\cite{Peeb80}).
When clustering is present, these errors will be different.
 
\section{Results}

The size and extent of the absorber catalog of Vanden Berk \etal\  (1996a)
permits exploration of clustering
over an unprecedented range in scale 
(from about 1 to over 300 \hMpc) 
{\it and} redshift ($z$ from 1.2 to 4.5).
Figure~1 shows the line--of--sight correlation function of \CIV\
absorbers, $\xi_{\rm aa}(\Delta r)$, as a function of absorber comoving 
separation, $\Delta r$, 
for the entire sample of absorbers.
The results are shown for both a $q_0=0.5$ ({\it top panel}, 25 \hMpc\ bins)
and a $q_0=0.1$ ({\it bottom panel}, 60 \hMpc\ bins) cosmology.\footnote
{Larger bins are required for $q_0=0.1$ because, at high redshift, a larger
comoving separation $\Delta r$ arises 
from a fixed redshift interval $\Delta z$.}
The vertical error bars through the data points are $1\, \sigma$ errors in the  
estimator for $\xi_{\rm aa}$,
which differ from the $1\, \sigma$ region of scatter ({\it dashed line},
calculated by Monte Carlo using the bootstrap technique described above)
around the no--clustering null hypothesis.

Remarkably, there appears to be significant clustering in 
the first four bins of Figure~1:
The mean correlation for those bins
is $\langle\xi_{\rm aa}\rangle = 0.42 \pm 0.10$ ($q_0=0.5$) or
$0.39 \pm 0.10$ ($q_0=0.1$).
Here the errors quoted are the $1\, \sigma$ errors in the estimate of the
correlation function (corresponding to the error bars in Figure 1).
In assessing the significance of these results, however,
the appropriate measure of departure from uniformity
is the $1\, \sigma$ region of scatter around the null hypothesis
(corresponding to the dashed lines in Figure 1),
which equals 0.085 ($q_0=0.5$) or 0.082 ($q_0=0.1$).
Thus, the positive correlation seen in the first four bins of
Figure~1 has a significance of $5.0\, \sigma\, (Q = 2.9\times 10^{-7})$, 
if $q_0=0.5$,
and $4.8\, \sigma\, (Q = 9.2\times 10^{-7})$ 
if $q_0=0.1$.
Therefore, there is significant evidence of clustering of matter
traced by \CIV\ absorbers on scales up to 100 \hMpc\ ($q_0=0.5$)
or 240 \hMpc\  ($q_0=0.1$).

There is no evidence from Figure~1
for clustering on comoving scales greater than these.
The mean correlation of the last six bins in Figure~1 
is $\langle\xi_{\rm aa}\rangle = 0.01 \pm 0.08 $ ($q_0=0.5$ or 0.1),
which is consistent with zero.

We have investigated the evolution of the superclustering by dividing
the absorber sample into three approximately equal redshift sub--samples.
Figure~2 shows $\xi_{\rm aa}(\Delta r)$,
as a function of absorber comoving separation, $\Delta r$ ($q_0=0.5$),
for low ($1.2<z<2.0$, {\it top panel}), 
medium ($2.0<z<2.8$, {\it middle panel}), 
and high ($2.8<z<4.5$, {\it bottom panel})
redshift absorber sub--samples.
Larger bins for the comoving separation were used in Figure~2
because of the smaller number of absorbers in each sub--sample.
The error bars are as in Figure~1.

The correlation function estimator is positive in the
first two bins (corresponding to scales of 100 \hMpc\ and less, 
like the first four
bins in Figure~1, {\it top panel})
in all three panels of Figure~2:
The mean correlation ($q_0=0.5$)
is $\langle\xi_{\rm aa}\rangle = 
0.40 \pm 0.17\, (2.7\, \sigma)$
for the low redshift sub--sample, 
$\langle\xi_{\rm aa}\rangle = 0.32 \pm 0.14\, (2.4\, \sigma)$
for the medium redshift sub--sample, 
and $\langle\xi_{\rm aa}\rangle = 0.72 \pm 0.25\, (3.4\, \sigma)$ 
for the high redshift sub--sample.
The values for $q_0=0.1$ are essentially the same as these.

Thus, the significant superclustering seen in Figure~1 is
present in all three  redshift sub--samples in Figure~2,
so that the superclustering  is present
even at redshift $z \gtrsim 3$.
Furthermore, it does not appear that the superclustering scale,
in comoving coordinates,
has changed significantly since then.

\section{Discussion}

We have found evidence for the existence of
large--scale superclustering
that has existed
since redshift $z\gtrsim 3$ and
has a comoving scale $\sim 100$ \hMpc\ ($q_0=0.5$)
or 240 \hMpc\ ($q_0=0.1$) that has not changed much since then.
We have examined the clustering signal more closely
and find that a large portion comes from
7 QSO lines of sight that have groups of 
4 or more \CIV\ absorbers 
within a 100 \hMpc\ interval ($q_0=0.5$).
(From Monte Carlo simulations,
we expect only $2.7\pm 1.5$ QSOs with such groups.)
When these are removed, the mean correlation function on
scales of 100 \hMpc\ and less ($q_0=0.5$) is
$\langle\xi_{\rm aa}\rangle = 0.16 \pm 0.14$.
Table 1 lists these 7 QSOs, along with the properties of their groups
of \CIV\ absorbers.

Several of the groups in Table 1 have been previously identified as
potential superclusters (\cite{Rom91,AS94});
indeed, superclustering has been suggested 
because of an unusual concentration of absorbers,
within several tens of megaparsecs,
in the lines of sight
of 1037--2704 and 1038--2712
(\cite{Jakob86,SS87,R87,DI96,Petit96}),
and 0237--233 (\cite{Boisse87,HHW89,Foltz93}).
However, there was concern that any inferred
superclustering might be overestimated because of aliasing
of small--scale power.
Like Dinshaw \& Impey (1996),
we find that superclustering is {\it not} caused by aliasing,
since it remains after combining all absorbers lying within
3.5 \hMpc\ of each other into a single system.
(For example, we count the 11 absorbers of 0237--233 in a 100 \hMpc\
interval as being 4 independent systems.)
We have found two potential superclusters in the spectra of QSOs
2126--158 and 2359+068, at redshift $z\sim 2.8$.

The superclustering is
indicative of generic large--scale clustering
in the universe, out to high redshift $z \gtrsim 3$,
on a scale frozen in comoving coordinates 
that is --- if $q_0=0.5$ --- similar to the size of the voids and walls in
galaxy redshift surveys of the local
universe ($z < 0.2$), such as the CfA survey
(\cite{GH89})  and others (\cite{Kir81,Chin83,daCosta94,Landy96}).
It also appears consistent with the general finding of
Broadhurst \etal\ (1990) that galaxies are clustered
on very large scales,
although we have not confirmed that  there is 
quasi--periodic clustering with power
peaked at 128 \hMpc .
It may also have been found in quasars (\cite{Krav96}).

Our estimate of the superclustering scale increases to 240 \hMpc\ if
$q_0=0.1$ (see Figure~1),
which is larger than the largest scales of clustering known at present.
If the structures traced by
\CIV\ absorbers are of the same nature as those seen locally
in galaxy redshift surveys,
the superclustering scale should have
a value closer to 100 \hMpc\ .
This may be indicative of a larger value of $q_0$,  and hence $\Omega_0$.

In Figure~3, we show the 
mean correlation function, $\langle\xi_{\rm aa}\rangle$,
on scales of 50 \hMpc\ and less, as a function of redshift $z$,
for the entire data set of absorbers ({\it solid symbols}),
and for strong \MgII\ absorbers (\cite{SS92}, {\it open symbol}),
with $q_0=0.5$.
Also shown is its expected dependence on redshift, 
calculated from the growth factor (\cite{Peeb80}, 1993),
normalized to the sample mean correlation of 0.48
at the sample median redshift of $z=2.31$, 
for $\Omega_0 =1$ ({\it solid line})
and $\Omega_0=0.1$ ({\it dashed line}).
While the evolution of the correlation function 
on 50 \hMpc\ scales is consistent
with that expected in cosmologies with
density parameter ranging from $\Omega_0= $ 0.1 to 1,
the Sloan Digital Sky Survey
should give better estimates for $\langle\xi_{\rm aa}\rangle$
(see the typical error bar at the upper right of Figure~3)
and will be able to constrain $\Omega_0$ to 20\%.

It is possible that structures of size $\sim 100$ \hMpc\
may have arisen through the  Zel'dovich ``pancake'' scenario
(\cite{Zel70,Mel90,Peebles93}),
and that this particular scale has deep significance because
it is of order the horizon size at decoupling,
and is the scale on which an ``acoustic peak'' is expected in the initial
power spectrum of many cosmologies (\cite{Peebles93,Landy96}).
The lack of clustering on scales greater than 100 \hMpc\
has implications on the peak and turn--over scale of the power
spectrum (\cite{Quash96}).

\acknowledgments
We acknowledge helpful discussions with
Don Lamb, Paul Ricker, Bob Rosner, and Alex Szalay, 
and statistical comments by Carlo Graziani.
Damian Bruni and Chris Mallouris have greatly assisted in compiling
the new catalog.
JMQ is supported by the Compton Fellowship -- NASA grant GDP93-08.
DEVB was supported in part by the Adler Fellowship at the University of
Chicago, and by NASA Space Telescope grant GO-06007.01-94A.  

\clearpage

\begin{deluxetable}{cccccc}

\tablecaption{QSOs with 4 or more \CIV\ absorbers 
in a 100 \hMpc\ interval ($q_0=0.5$). \label{tbl-1}}

\tablehead{
\colhead{QSO} & \colhead{$N_{\rm exp}$} & \colhead{$N_{\rm abs}$} 
& \colhead{$\langle z_{\rm abs}\rangle$} &
\colhead{$\Delta z_{\rm abs}$} & \colhead {$\Delta r$ (\hMpc)}
}
\startdata
0149+336  & 0.57 & 4 & 2.1583 & 0.1593 & 85.8 \nl
0237--233 & 0.30 & 4 & 1.6348 & 0.0775 & 54.4 \nl
0958+551  & 0.25 & 4 & 1.3207 & 0.1045 & 88.5 \nl
1037--270 & 0.90 & 5 & 2.0133 & 0.1703 & 98.5 \nl
1038--272 & 0.79 & 4 & 1.9283 & 0.1642 & 98.2 \nl
2126--158 & 0.47 & 5 & 2.7265 & 0.1814 & 75.6 \nl
2126--158\tablenotemark{\ast} & 0.74 & 5 & 2.7804 & 0.2280 & 92.6 \nl
2359+068  & 0.51 & 6 & 2.8303 & 0.2125 & 84.8 \nl
2359+068\tablenotemark{\ast}  & 0.64 & 6 & 2.8730 & 0.2394 & 94.5 \nl

\enddata

\tablenotetext{\ast}{Part of this group forms another group (also listed)
with an additional absorber within the same QSO, 
but the extent of the two groups taken together exceeds 100 \hMpc .
$N_{\rm exp}$ is the number of expected absorbers in the interval 
$\Delta z_{\rm abs}$.}

\end{deluxetable}

\newpage
\figcaption[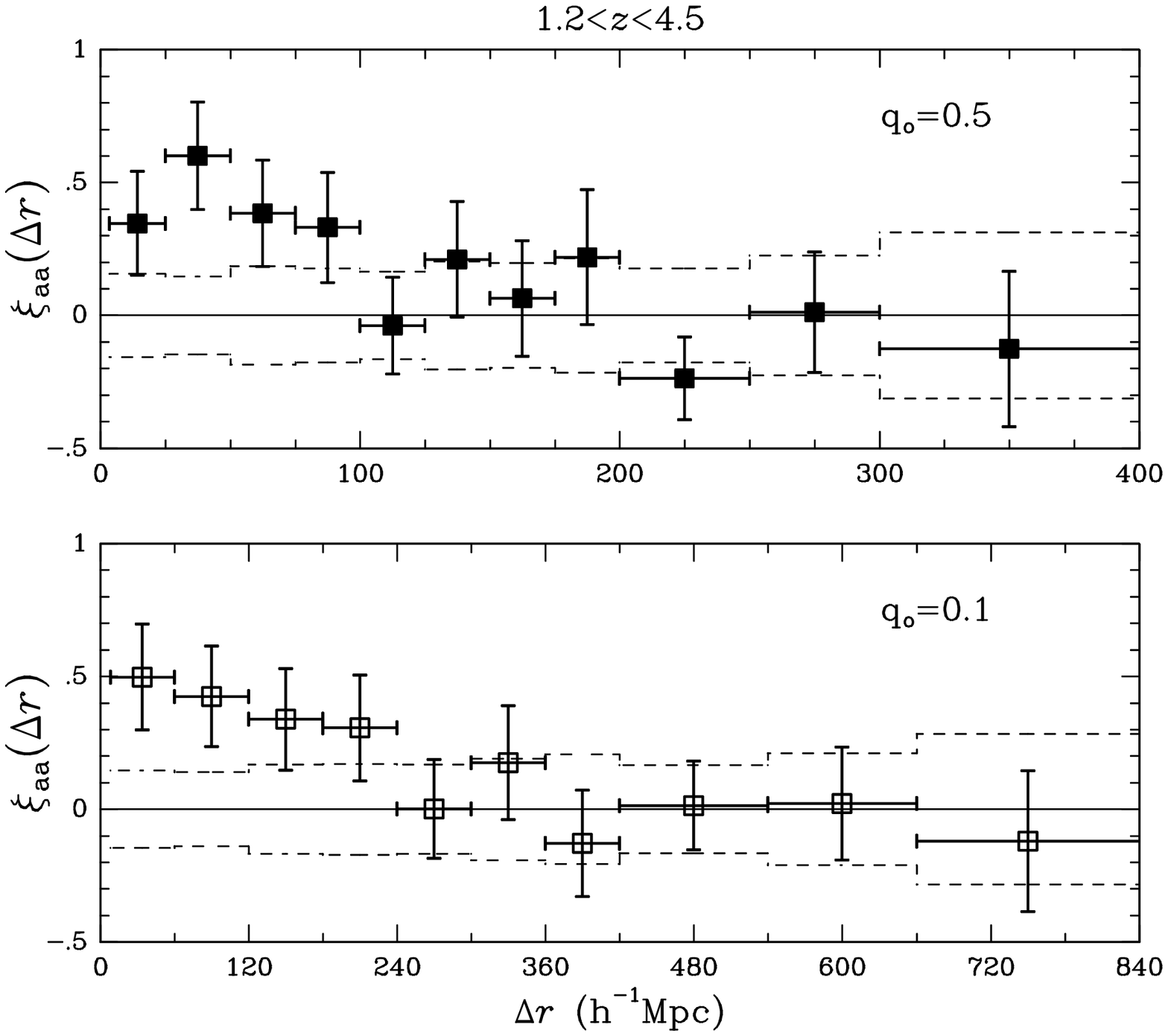]{Line--of--sight correlation function of \CIV\
absorbers, $\xi_{\rm aa}(\Delta r)$, as a function of 
absorber comoving separation, $\Delta r$, 
for the entire sample of absorbers,
in a $q_0=0.5$ ({\it top panel}, 25 \hMpc\ bins)
and a $q_0=0.1$ ({\it bottom panel}, 60 \hMpc\ bins) cosmology.
The vertical error bars through the data points 
are $1\, \sigma$ errors in the  
estimator for $\xi_{\rm aa}$,
which differ from the
$1\, \sigma$ region of scatter ({\it dashed line}) around the
null hypothesis of no clustering ($\xi_{\rm aa}=0$). \label{fig1}}

\figcaption[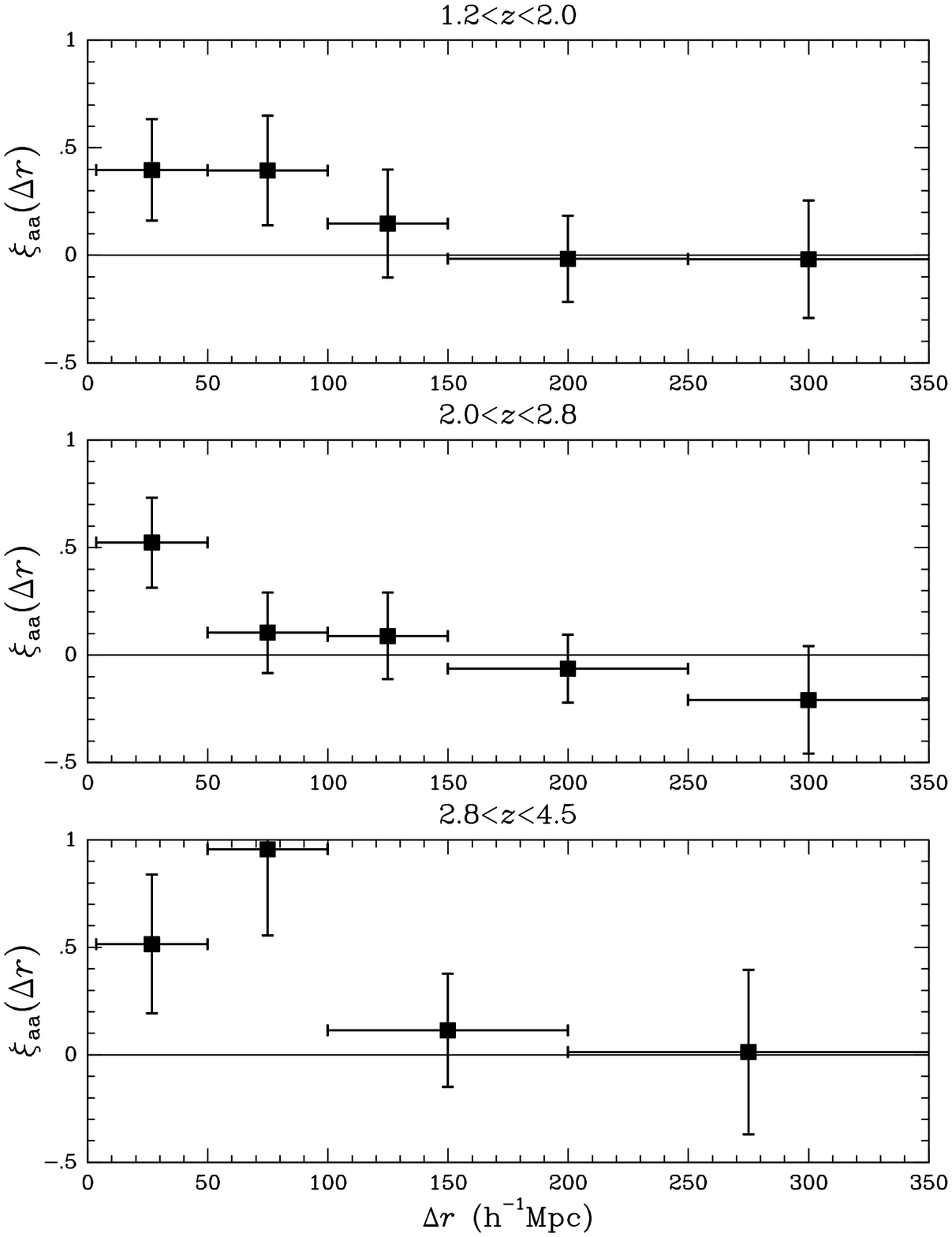]{Line--of--sight correlation function of \CIV\
absorbers, $\xi_{\rm aa}(\Delta r)$, as a function of
absorber comoving separation, $\Delta r$
($q_0=0.5$),
for low ($1.2<z<2.0$, {\it top panel}), 
medium ($2.0<z<2.8$, {\it middle panel}), 
and high ($2.8<z<4.5$, {\it bottom panel}) 
redshift absorber sub--samples .
The vertical error bars through the data points
are $1\, \sigma$ errors in the  
estimator for $\xi_{\rm aa}$. \label{fig2}}

\figcaption[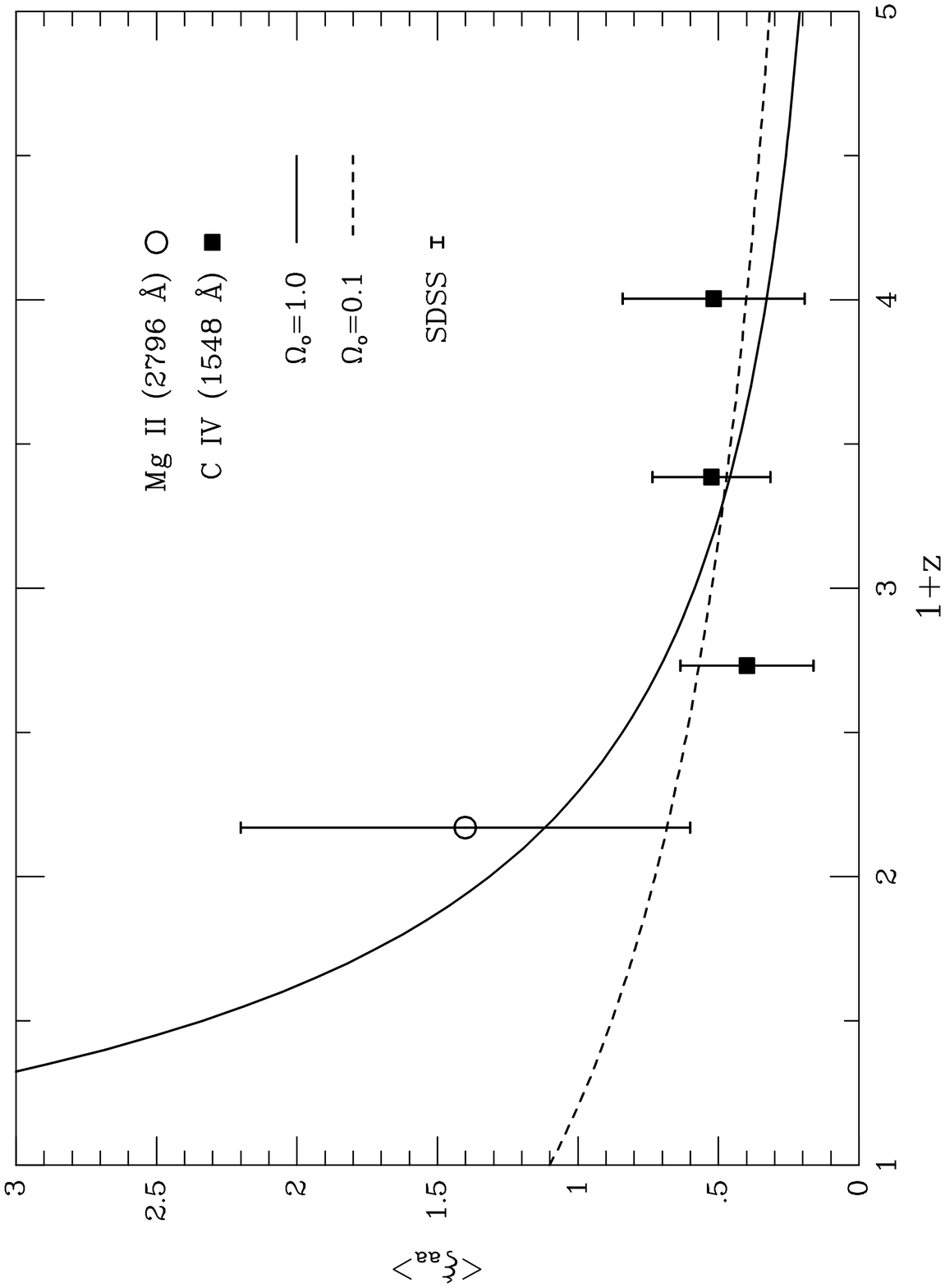]{Mean correlation function,
$\langle\xi_{\rm aa}\rangle$, on scales of 50 \hMpc\ and less,
as a function of redshift $z$, for the entire data set of
\CIV\ absorbers ({\it solid symbols}), and for strong \MgII\
absorbers (\protect\cite{SS92},
{\it open symbol}), with $q_0=0.5$.
Also shown is its expected dependence on redshift,  
normalized to the sample mean of 0.48 at the median sample redshift of
$z=2.31$,  for $\Omega_0 =1$ ({\it solid line}) and
$\Omega_0=0.1$ ({\it dashed line}). The expected error bar
from the Sloan Digital Sky Survey is shown in the upper right.
\label{fig3}}

\newpage

\begin{figure}
\figurenum{1}
\plotone{fig1.eps}
\caption{}
\end{figure}

\newpage
\begin{figure}
\figurenum{2}
\plotone{fig2.eps}
\caption{}
\end{figure}

\newpage
\begin{figure}
\figurenum{3}
\plotone{fig3.eps}
\caption{}
\end{figure}

\end{document}